 \documentclass{article}
 \usepackage{graphicx}
 \usepackage{natbib}

\usepackage{amsmath}
\usepackage{amssymb}
\usepackage[figuresright]{rotating}
\usepackage[colorlinks,citecolor=blue,urlcolor=blue,filecolor=blue,backref=page]{hyperref}
\usepackage{url}
\usepackage[normalem]{ulem}

 \usepackage{enumerate}

\newcommand{\ba}{ {\boldsymbol a} }
\newcommand{\bA}{ {\boldsymbol A} }

\newcommand{\bB}{ {\boldsymbol B} }

\newcommand{\bW}{ {\boldsymbol W} }

\newcommand{\bX}{ {\boldsymbol X} }

\newcommand{\bY}{ {\boldsymbol Y} }

\newcommand{\bbeta}{ {\boldsymbol \beta} }

\newcommand{\bet}{ {\boldsymbol \eta} }

\newcommand{\bsigma}{ {\boldsymbol \sigma} }

\newcommand{\bSigma}{ {\boldsymbol \Sigma} }

\title{Bayesian Mixed Effect Sparse Tensor Response Regression Model with Joint Estimation of Activation and Connectivity}
\author{Daniel Spencer, Rajarshi Guhaniyogi, Raquel Prado \\
	Department of Statistics, Baskin School of Engineering, \\ UC Santa Cruz, Santa Cruz, CA 95064}

\begin{document}

 \maketitle

\begin{abstract}
Brain activation and connectivity analyses in task-based functional magnetic resonance imaging (fMRI) experiments with multiple subjects are currently at the forefront of data-driven neuroscience.  In such experiments, interest often lies in understanding activation of brain voxels due to external stimuli and strong association or connectivity between the measurements on a set of pre-specified group of brain voxels, also known as regions of interest (ROI). This article proposes a joint Bayesian additive mixed modeling framework that simultaneously assesses brain activation and connectivity patterns from multiple subjects. In particular, fMRI measurements from each individual obtained in the form of a multi-dimensional array/tensor at each time are regressed on functions of the stimuli. We impose a low-rank PARAFAC decomposition on the tensor regression coefficients corresponding to the stimuli to achieve parsimony. Multiway stick breaking shrinkage priors are employed to infer activation patterns and associated uncertainties in each voxel. Further, the model introduces region specific random effects which are jointly modeled with a Bayesian Gaussian graphical prior to account for the connectivity among pairs of ROIs. Empirical investigations under various simulation studies demonstrate the effectiveness of the method as a tool to simultaneously assess brain activation and connectivity. The method is then applied to a multi-subject fMRI dataset from a balloon-analog risk-taking experiment in order to make inference about how the brain processes risk.

\end{abstract}

\section{Introduction}
\label{sec:intro}

Of late, rapid advancements in different imaging modalities have generated massive neuroimaging data which are key in understanding how the human brain functions. For the present article, our motivation is mainly drawn from multi-subject functional MRI (fMRI) studies. In the context of an fMRI scan, the brain at a single point in time is envisioned as a three-dimensional tensor partitioned into small cubes, known as \textit{voxels} \citep{lazar2008statistical}. A relative measure of oxygen in the blood, referred to as the blood oxygen level-dependent (BOLD) measure is obtained from every voxel in each scan typically acquired around every two seconds. Such scans can be taken when a subject is in a resting state without any conditions imposed; when a subject is exposed to certain conditions/stimuli, such as noises or videos; or when a subject is actively involved in completing a task. As oxygen is required to perform functions in the brain, these readings are used to infer which parts of the brain are ``active" in a given thought process. Expected activation patterns include local spatial dependence in the sense that voxels located next to each other tend to be jointly activated, as well as non-local dependencies in which groups of voxels in distant regions of the brain are activated by a given thought process. 

In addition to determining brain activation linked with a given cognitive or sensorimotor function, neuroscientists are often interested in the way different spatially-adjacent groups of voxels, referred to as Regions of Interest (ROIs) in the brain work together to process information. These types of relationships between ROIs are collectively referred to as \textit{functional connectivity} \citep{hutchison2013dynamic}. The major contribution of this article is the proposal of a Bayesian modeling framework that simultaneously detects voxel level activation and connectivity between different ROIs with precise characterization of uncertainty for multi-subject fMRI data. Simultaneous analysis of multi-subject 3D fMRI scans is a challenging problem due to the sheer amount of data.  Our modeling framework addresses this issue by a mixed-effects tensor response regression analysis in which low-rank tensor decompositions are combined with a multiway stick-breaking shrinkage prior to achieve parsimony. Such framework provides a powerful and computationally-feasible setting for inferring activation and connectivity in multi-subject task-related fMRI studies.

Several approaches have been proposed for the analysis of brain activation. Single-subject frameworks in particular have a rich background for modeling activation. The simplest of them fits a regression model at each voxel with the observed voxel-specific fMRI measure as response regressed on activation related predictors, and identifies if the response is significantly associated with the predictors in that voxel \citep{friston1995spatial,penny2011statistical}, after accounting for multiple correction. Although conceptually simple, this approach fails to accommodate spatial association across voxels. Another idea, which addresses the sparse nature of fMRI activation, assigns the spike-and-slab prior on the regression coefficients \citep{brown1998multivariate, yu2018bayesian} across all voxels. Various approaches also account for spatial association in the neighboring voxels by inducing dependence among voxel-specific regression coefficients using Markov Random Fields \citep{zhang2014spatio,kalus2014classification,smith2007spatial,lee2014spatial}. \citet{guhaniyogi2017bayesian} proposes a tensor decomposition method with shrinkage priors to model activation in the setting with a scalar response and a tensor predictor. \cite{guhaniyogispencerbayesian} extends this to a tensor-valued response and a scalar predictor without accounting for any inter-regional connectivity. This approach focuses substantially on the posterior contraction theory of tensor response regression models with a brief simulation study and real data analysis illustrating the approach merely for single-subject fMRI analysis. Other sophisticated approaches include spatially varying coefficient (SVC) models which employ spatial basis functions to model activation-related coefficients \citep{flandin2007bayesian,zhu2014spatially}. Besides being computationally expensive, such models are sensitive to the selection of the basis functions, and require specific knowledge to appropriately calibrate them. More recently, a class of approaches proposes joint modeling of BOLD measures across all voxels in the form of a tensor response \citep{li2017parsimonious}. Though potentially useful for inferring activation, these approaches have not been developed for multi-subject studies. 

In order to overcome the computational challenge of having voxel-level data analyzed for multiple subjects, early approaches combined information across voxels, either using generalized linear model (GLM) parameter estimates or residual variance. Two-stage methods by \citet{bowman2008bayesian} and \citet{sanyal2012bayesian} fitted subject-specific GLMs, and then used regularization on the parameter estimates to determine activation. Mixture models and non-parametric Bayesian models have also been proposed to analyze inter- and intra-subject variability, though they incur heavy computational cost. 


While there is considerable literature on activation-only models, literature on Bayesian functional connectivity has witnessed a few very distinct approaches. To this end, \citet{patel2006bayesian} and \citet{patel2006determining} discretized the fMRI time series between regions based on whether they had elevated activity according to a threshold, and then compared joint and marginal probabilities of elevated activity. \citet{bowman2008bayesian} modeled similarity within and between regions of interest based on estimates of elements of the covariance in a two-stage model. A Dirichlet process prior was used by \citet{zhang2014spatio} to cluster remote voxels together, asserting that the clustering inferred an inherent connectivity. \citet{zhang2014inferring} went on to propose a dynamic functional connectivity model, estimating connective phases and temporal transitions between them.

As mentioned above, there are several Bayesian modeling frameworks for assessing activation or connectivity separately, however, models incorporating both of them jointly in multi-subject fMRI studies with voxel-level data are comparatively rare in the literature. In the recent past, \citet{kook2017npbayes} proposes such a model in which a Dirichlet process (DP) mixture model is used to classify voxels as active via discrete wavelet transformations. The clustering of the voxels through time via the mixture components is then used to derive a measure of inter-voxel connectivity within- and between-subjects. While their model succinctly captures activation and connectivity, the use of a Dirichlet process may hinder computational efficiency. Variational Bayes methods were used to speed-up computation, which provide approximate posterior results in a fraction of the time that a full Markov Chain Monte Carlo simulation would require.


Our article proposes a multi-subject Bayesian tensor mixed effect model that estimates voxel-wise activation and inter-regional connectivity simultaneously through a novel adaptation of shrinkage priors. To elaborate further, the model envisions BOLD measures over all voxels together for a subject at any time as a tensor response and regresses this tensor object on the activation related predictor.
In order to achieve substantial parsimony, the coefficient tensor is assumed to possess a low-rank PARAFAC decomposition, and a multiway stick-breaking shrinkage prior is assigned on the tensor coefficient to shrink the cells corresponding to unimportant voxels close to zero, while maintaining accurate estimation and uncertainty of cell coefficients related to important voxels.  One of the main advantages of using tensor representations is that these are able to capture local and non-local spatial effects without explicitly introducing a spatial structure into the model. This is an advantage with respect to approaches such as those in \cite{kook2017npbayes} which model local spatial correlation among voxels using Markov random field priors. In addition, our proposed model incorporates subject-ROI-specific random effects with a Gaussian graphical prior, imposing regularization on the precision matrix of the effects between regions \citep{wang2014regularized}. Both the activation and connectivity parameters are then classified into zero- and nonzero-effect sizes using the sequential 2-means method proposed by \citet{li2017variable}. As a result, the model produces accurate measures of voxel-wise activation and inter-regional connectivity with interpretable effect sizes and uncertainty quantification without the need for fine-tuning hyperparameters or basis functions. In addition, the model is scalable and computationally-efficient enough to provide samples from the exact posterior distribution for 2-D slices or 3-D volumes of brain images, as well as higher-order tensor images.

The upcoming sections proceed as follows. The model, including the prior structure, is set forth in Section \ref{sec:methodology}. 
 Section \ref{sec:post} discusses posterior inference.
 Section \ref{sec:simulation_studies} empirically validates the model with simulation studies. Sensitivity to hyperparameter specification is also demonstrated to illustrate the robustness of the model. Section \ref{sec:real_data_analysis} describes the multi-subject fMRI data from the balloon-analog risk-taking experiment in detail. Finally, Section \ref{sec:discussion} presents a discussion of the results and describes some future extensions.

\section{Methodology}
\label{sec:methodology}

This section begins by setting the tensor notation and the PARAFAC decomposition structure. The complete modeling framework, including prior structure and hyperparameter specification, is then detailed. 

\subsection{Notation and preliminaries}\label{sec:notations}
A tensor $\bA$ is a $D-$way array (also known as a $D$-th order tensor) of dimensions $p_1\times \cdots\times p_D$ with 
$(i_1,...,i_D)-th$ cell entry denoted by $A[i_1,...,i_D]\in\mathbb{R}$, $i_1=1,..,p_1; \ldots; i_D=1,...,p_D$. For vectors $\ba_1$,...,$\ba_D$ of lengths $p_1,...,p_D$ respectively, define the outer product between the vectors, denoted by
$\bA=\ba_1\circ\cdots\circ\ba_D$, as a $D-$way array with $(i_1,...,i_D)-th$
cell element $A[i_1,...,i_D]=\prod_{j=1}^{D}a_{j,i_j}$, where $a_{j,i_j}$ is the $i_j-th$ element of $\ba_j$. $\bA$ is referred to as a rank 1 tensor with dimensions $p_1\times\cdots\times p_D.$ A rank-R tensor $\bA$ is obtained by summing $R$ rank 1 tensors,
$\bA=\sum_{r=1}^R \ba_{1,r}\circ\cdots\circ\ba_{D,r}$. This is also referred to as the CP/PARAFAC decomposition of rank $R$ \citep{kiers2000towards} and is used due to its relative simplicity \citep{kolda2009tensor}. In what follows, we will refer to $\ba_{j,r}$'s as margins of the tensor $\bA$.

\subsection{Model framework and prior structure}
 Let $\bY_{i,g,t}$ be the observed fMRI data in brain region $g$ for the $i$th subject at the $t$th time point. $\bY_{i,g,t}$ is observed in the form of a tensor of dimensions $p_{1,g}\times\cdots\times p_{D,g}$. In the context of fMRI data analysis, $D$ is two or three, depending on whether a single slice or the entire brain volume is analyzed. To simultaneously measure activation due to stimulus at voxels in the $g$th brain region and connectivity among $G$ brain regions, we employ an additive mixed effect model with tensor-valued fMRI response and activation related predictor $x_{i,t}\in\mathbb{R}$,
\begin{equation}
\mathbf{Y}_{i,g,t} = \mathbf{B}_g x_{i,t} + d_{i,g} + \mathbf{E}_{i,g,t},
\label{eg:lik_tensor}
\end{equation}
for subject $i = 1,\ldots,n$, in region of interest $g = 1,\ldots,G$, and time $t = 1,\ldots,T$. Elements in the error tensor $\mathbf{E}_{i,g,t}$ are assumed to be normally distributed with mean $0$ and shared variance $\sigma_y^2$, though our framework can be extended to incorporate temporally correlated errors. Without loss of generality, the response tensor $\mathbf{Y}_{i,g,t}$ in the proposed model is centered over each ROI to eliminate the need for an intercept term.

The tensor coefficient $\mathbf{B}_g \in \mathbb{R}^{p_{1,g}\times\cdots\times p_{D,g}}$ is used to infer the strength of the association between $x_{i,t}$ and each voxel in $\mathbf{Y}_{i,g,t}$. In particular, $B_g[i_1,...,i_D]=0$ implies that the $(i_1,...,i_D)$th voxel in the $g$th ROI is \emph{not activated} by the stimulus. In fact, the activation pattern is typically sparse and localized with only a few nonzero elements in $\bB_g$. $d_{i,g} \in \mathbb{R}$ are region- and subject-specific random effects which are jointly modeled to borrow information across ROIs. In the present context, the conditional distributions $(d_{i,g},d_{i,g'})|\{d_{i,g''}:g''\neq g,g'\}$ are investigated to assess the strength of connectivity between a pair of regions. As part of the model development, we impose prior distributions that favor conditional independence between most pairs $d_{i,g}$ and $d_{i,g'}$, estimating connectivity only among a few pairs of regions.

 As mentioned above, the coefficient tensor $\mathbf{B}_g \in \mathbb{R}^{p_{1,g} \times \cdots \times p_{D,g}}$ in equation (\ref{eg:lik_tensor}) characterizes a sparse relationship between the tensor response and the time-varying covariate $x_{i,t}$ in region $g$. In order to achieve parsimony in the number of estimated parameters, $\mathbf{B}_g$ is assumed to have a rank $R$ PARAFAC decomposition.

\begin{align}\label{PARAFAC}
\mathbf{B}_g & = \sum_{r=1}^R \boldsymbol{\beta}_{g,1,r} \circ \cdots \circ \boldsymbol{\beta}_{g,D,r},
\end{align}
with tensor margins $\bbeta_{g,1,r},..,\bbeta_{g,D,r}.$ The PARAFAC tensor decomposition dramatically reduces the number of parameters in $\bB_g$ from $\prod_{j=1}^D p_{j,g}$ to $R\sum_{j=1}^{D} p_{j,g},$ with the extent of the achieved parsimony being dependent on $R$. Note that a smaller value of $R$ leads to parsimony and computational gain, perhaps at the cost of inferential accuracy. In contrast, a choice of even moderately large $R$ entails higher computation cost. Again, using $R$ as a model parameter often increases computation cost and is deemed unnecessary \citep{guhaniyogi2017bayesian}. In view of the earlier literature, this article proposes fitting the model with various choices of $R$ and chooses the one that yields the lowest Deviance Information Criterion (DIC) \citep{gelman2014bayesian}. More discussion on the choice of $R$ is provided in Section~\ref{sec:simulation_studies}.

A critical question remains how to devise a prior distribution on the low-rank decomposition (\ref{PARAFAC}) to facilitate identifying geometric sub-regions in the tensor response which share an association with the predictor. Additionally, the model intends to build joint priors on region specific random effects $d_{i,g}$s to assess connectivity patterns. The next two sections propose careful elicitation of the prior distributions on $\bB_g$ and $d_{i,g}$ to achieve our stated goals.
\subsection{Multiway stick breaking shrinkage prior on $\bB_g$ to assess activation}\label{sec:activation}
Although the spike-and-slab priors for selective predictor inclusion \citep{george1993variable,ishwaran2005spike} possess attractive theoretical properties and an easy interpretation, they often lose their appeal due to their inability to explore a large parameter space. As a computationally-convenient alternative, an impressive variety of shrinkage priors  \citep{carvalho2010horseshoe,armagan2013generalized} in the context of ordinary Bayesian high dimensional regression have been developed. 

Shrinkage architecture relies on shrinking coefficients corresponding to unimportant predictors, while maintaining accurate estimation with uncertainty for important predictor coefficients. The existing shrinkage prior literature serves as a basis to the development of shrinkage priors on the tensor coefficients. However, constructing such a prior on $\bB_g$ presents additional challenges. To elaborate on it, notice that proposing a prior on a low-rank PARAFAC decomposition of $\bB_g$ is equivalent to specifying priors over tensor margins $\bbeta_{g,j,r}$. Since every cell coefficient in $\bB_g$ is a nonlinear function of the tensor margins, careful construction of shrinkage priors on $\bbeta_{g,j,r}$s is important to impose desirable tail behavior of $B_g[i_1,...,i_D]$ parameters. 
To this end, this article employs \emph{a multiway stick-breaking} shrinkage prior on $\bB_g$ to ensure desirable tail behavior. More specifically, the following shrinkage prior is proposed on the 
tensor margins
\begin{align*}
\boldsymbol\beta_{g,j,r} \sim  \text{N}\left(\mathbf{0},\phi_{g,r}\tau_g\mathbf{W}_{g,j,r}\right),\: \mathbf{W}_{g,j,r} = \text{diag} \left(\omega_{g,j,r,1},\ldots,\omega_{g,j,r,p_j}\right),
\end{align*}
where
\begin{align*}
\omega_{g,j,r,\ell}  \sim  \text{Exp} \left( \frac{\lambda_{g,j,r}^2}{2} \right) ,\:
\lambda_{g,j,r} \stackrel{iid}{\sim} \text{Gamma}(a_\lambda,b_\lambda),
\end{align*}
for $j=1,...,D$ and $g=1,...,G$.
This prior defines a set of rank specific scale parameters $\phi_{g,r}$ using a stick breaking construction of the form $\phi_{g,r}=\xi_{g,r}\prod\limits_{l=1}^{r-1}(1-\xi_{g,l})$, $r=1,...,R-1$, and
$\phi_{g,R}=1-\sum\limits_{r=1}^{R-1}\phi_{g,r}=\prod\limits_{l=1}^{R-1}(1-\xi_{g,l})$ that achieves efficient shrinkage across ranks, where $\xi_{g,r}\stackrel{iid}{\sim}Beta(1,\alpha_g)$. The global scale parameters are modeled as $\tau_1,...,\tau_G\stackrel{iid}{\sim} IG(a_{\tau},b_{\tau}).$  

Flexibility in modeling tensor margins are accommodated by introducing $\bW_{g,j,r}$s. In fact, integrating out $\bW_{g,j,r}$ and $\lambda_{g,j,r}$ yields a generalized double Pareto shrinkage prior for the elements of $\bbeta_{g,j,r}$ conditional on $\phi_{g,r}$ and $\alpha_g$.

Without constraints on the values for $\phi_{g,r} \in \boldsymbol{\Phi}_g$, where $r = 1,\ldots,R$, identifiability issues arise in the posterior sampling for the variance terms for $\bbeta_{g,j,r} \in \bB_g$. In order to address this issue, a stick-breaking structure is imposed on $\phi_{g,r}$'s, as described above. In effect, this prevents $\phi_{g,r}$'s from switching labels across ranks in which the variance of $\bbeta_{g,j,r}$ may be close together. The result of this constraint is a more stable MCMC for the posterior draws of $\bbeta_{g,j,r}$. The tuning parameter $\alpha_g$ in the stick-breaking construction assumes a crucial role in determining which tensor rank $R$ is favored by data. In particular, $\alpha_g\rightarrow 0$ favors small values of most $\phi_{g,r}$ a-priori. Therefore, a data-dependent learning of $\alpha_g$ is essential in order to tune to the desired sparsity in $\bB_g$. Section~\ref{subsec:hyperspec} discusses a model-based choice of $\alpha_g$, along with the specific choices for $a_{\lambda},b_{\lambda},a_{\tau}$, and $b_{\tau}$.
\subsection{Bayesian Graphical Lasso Prior for modeling connectivity}\label{sec:connectivity}

Following \cite{wang2012bayesian}, to capture connectivity between different regions for individuals, $d_{i,g}$s are jointly modeled with a Gaussian graphical lasso prior. To be more precise, 
\begin{align}\label{graph_lasso}
&\mathbf{d}_i=(d_{i,1},..,d_{i,G})'\sim \text{N} (\mathbf{0},\boldsymbol{\Sigma}^{-1}),\:\:i=1,...,n\nonumber\\
& p(\bsigma|\zeta)=C^{-1}\prod_{k<l}\left[DE(\sigma_{kl}|\zeta)\right]
\prod_{k=1}^G\left[\text{Exp}(\sigma_{kk}|\frac{\zeta}{2})\right]{\boldsymbol 1}_{\bSigma\in\mathcal{P}^{+}},
\end{align}
where $\mathcal{P}^{+}$ is the class of all symmetric positive definite matrices and $C$ is a normalization constant. $\bsigma=(\sigma_{kl}:k\leq l)$ is a vector of upper triangular and diagonal entries of the precision matrix $\bSigma$. Using properties of multivariate Gaussian distribution, a small value of $\sigma_{kl}$ stands for weak connectivity between ROIs $k$ and $l$, given the other ROIs. In fact, $\sigma_{kl}=0$ ($k<l$) implies that there is no connectivity between ROIs $k$ and $l$, given the other ROIs. Thus, to favor shrinkage among off-diagonal entries of $\bSigma$ for drawing inference on connectivity between pairs of ROIs, the Bayesian graphical lasso prior employs double exponential prior distributions on the off-diagonal entries of this precision matrix. The diagonals of $\bSigma$ are assigned exponential distributions. Let $\bet=(\eta_{kl}:k<l)$ be a set of latent scale parameters. Using the popular scale mixture representation of double exponential distributions \citep{wang2012bayesian}, we
can write 
$$ p(\bsigma| \zeta) = \int p(\bsigma| \zeta, \bet) p(\bet| \zeta) d \bet,$$ with $p(\bsigma|\zeta,\bet)$ given by
\begin{align}
p(\bsigma|\zeta,\bet)=C_{\bet}^{-1}\prod_{k<l}\left[\frac{1}{\sqrt{2\pi\eta_{kl}}}\exp\left(-\frac{\sigma_{kl}^2}{2\eta_{kl}}\right)\right]
\prod_{k=1}^G\left[\frac{\zeta}{2}\text{Exp}(-\frac{\zeta}{2}\sigma_{kk})\right]{\boldsymbol 1}_{\bSigma\in\mathcal{P}^{+}},
\end{align}
where $C_{\bet}$ is the normalizing constant, which is an analytically intractable function of $\bet$. The mixing density of $\bet$ in the representation above is given by

\begin{align}
p(\bet|\zeta)\propto C_{\bet}\prod_{k<l}\frac{\zeta^2}{2}\exp(-\frac{\zeta^2}{2}\eta_{kl}).
\end{align}

The hierarchy is completed by adding a Gamma prior on $\zeta$, $\zeta\sim Gamma(a_{\zeta},b_{\zeta})$. Finally, an inverse gamma prior $\sigma_y^2 \sim \text{Inverse Gamma}(a_\sigma,b_\sigma)$ is used on the variance parameter $\sigma_y^2$. 






\subsection{Hyperparameter Specification}
\label{subsec:hyperspec}

The hyperparameters $\alpha_g$ in the stick-breaking construction play a key role in controlling the dimensionality of the model, with smaller values effectively favoring a low-rank tensor factorization. A discrete uniform prior is placed on the  $\alpha_g$ parameters over 10 equally-spaced values in the interval $[R^{-D},R^{-.10}]$, which will allow the data to dictate the level of sparsity appropriate for the prior \citep{guhaniyogi2017bayesian}. The posterior distribution of an $\alpha_g$ concentrated toward the left end of the interval encourages parsimony, while the posterior of an $\alpha_g$ concentrating at higher values permits a less sparse PARAFAC decomposition. The chosen prior range of $\alpha_g$ works for various simulation studies and moderate perturbation of the prior range seems to produce robust inference. The values chosen for $a_\lambda$ and $b_\lambda$ have a strong effect on the shrinkage properties of the generalized double-Pareto prior, and setting $a_\lambda = 3$ and $b_\lambda = \sqrt[2D]{a_\lambda}$ prevents the prior for $\lambda_{g,j,r}$ from allowing for insufficient variance for $B_g[i_1,...,i_D]$ to detect nonzero coefficients. Similar to \cite{guhaniyogi2017bayesian}, the hyperparameters $a_\tau$ and $b_\tau$ are set to $D - 1$ and $R^{1/D - 1}$, respectively, in order to prevent overshrinkage with higher tensor response dimensions. Following \citet{wang2014regularized}, $a_\zeta$ and $b_\zeta$ are set to 1 and 0.01, respectively, in order to preserve relative noninformativity of the Gaussian graphical prior. Finally, for both simulation studies and the real data analysis, $a_\sigma$ and $b_\sigma$ were set to be 1 and $-\log 0.95$, respectively. While these hyperparameters are specified to provide readers a specific set of choices and they produce desirable results, we establish in Section 4 that the inference is fairly robust with moderate perturbation of these hyperparameters.



\section{Posterior Computation}\label{sec:post}

The model framework and prior structure allow sampling from the posterior distribution using the Markov Chain Monte Carlo (MCMC) algorithm outlined in the supplementary material (Web Appendix A).
The posterior distributions of unknown quantities of interest are approximated by their empirical distributions from post burn-in MCMC samples.

Of particular interest in neuroscience is the assessment of whether a brain voxel is active or not, which, in our modeling framework translates to verifying whether $B_g[i_1,...,i_D]$ is nonzero for any voxel $(i_1,...,i_D)$. It is well-acknowledged that the problem of selecting important cell coefficients is a challenging task when $\bB_g$ is assigned a continuous shrinkage prior, since none of the cell coefficients is exactly zero in any MCMC iteration. Following the recent sequential 2-means variable selection approach of \cite{li2017variable}, we aim to address the problem of identifying significantly nonzero cell coefficients through post processing the posterior samples. The approach is based on first obtaining a posterior distribution of the number of signals by sequentially clustering the signal and the noise cell coefficients together, followed by estimating the signals from the posterior median. In the interest of space, we refer to Section 2.2 of \cite{li2017variable} for more details about the algorithm.

In order to obtain an interpretable measure the connectivity between regions, the partial correlation between regions is examined. Since the partial correlation accounts for the correlation between two regions after removing the influence of all other regions \citep{das2017interpretation}, it is expected to be the best measure of pairwise connections. First, the sequential two-means variable selection method \citep{li2017variable} was used on the posterior samples of the precision matrix $\boldsymbol{\Sigma}$ in order to select which precision elements were not equal to zero. The resulting precision matrix estimate was transformed into the partial correlation using the \texttt{prec2part} function in the \texttt{DensParcorr} package in R \citep{DensParcorr}. Regions with nonzero partial correlations are said to be connected \citep{warnick2018bayesian}.




\section{Simulation Studies}
\label{sec:simulation_studies}

To validate the proposed methods, we simulate synthetic data with similar structure to that found in data collected from human fMRI studies. The tensor responses are simulated considering a block experimental design from the likelihood in (\ref{eg:lik_tensor}). In each simulation study, we construct $G=10$ different coefficient tensors corresponding to disjoint spaces, hereafter referred to as regions. For ease of visualization, the coefficient tensors are created to be three-dimensional, but can be generalized to any arbitrary dimension $D$. Throughout the simulation study, a sample size of $n=20$ subjects is used, with the number of time points per subject being fixed at $T = 100.$ 

The covariate, $x_{i,t}=x_t$, was set to be the same for all of the subjects, without any loss of generality. A block experimental design is employed to generate the covariates, which consists of several discrete epochs of activity-rest periods, with the ``activity" representing a period of stimulus presentations, and the ``rest" referring to a state of rest or baseline. These activity-rest periods are alternated throughout the experiment to ensure that signal variation, scanner sensitivity and subject movement have the similar effect throughout the experiment. To simulate activity-rest periods, we use the stimulus indicator function $z_{t}$ as:

$$
z_{t} = 
    \begin{cases}
        1, & \text{ for } kP < t < kP + P/2, \quad k = 0,1,\ldots \\
        0, & \text{ otherwise}
    \end{cases}
$$
for all $t$, given a defined period $P$ for the block design. In our simulations, $P$ is set to be 30. Next, the \texttt{canonicalHRF} function in the \texttt{neuRosim} package in R \citep{welvaert2011neurosim} is used to convolve the stimulus indicator $z_{t}$ with the double-gamma canonical hemodynamic response function (HRF), which corrects for the expected delay between a stimulus and the resultant physiological response in the brain \citep{friston1998event}. This HRF is set using the default function values in \texttt{neuRosim} to have a delay of response relative to onset equal to 6 time steps, a delay of undershoot relative to onset of 12, a dispersion of response equal to 0.9, a dispersion of undershoot equal to 0.9, and a scale of undershoot equal to 0.35. The resulting covariate $x_{t}$ is plotted in Web Figure 1.
The dimensions of response tensor margins $p_{1,g},~p_{2,g},$ and $p_{3,g}$ are drawn from a Poisson distribution with a rate parameter of 10 for each region $g$. This generated tensor regions that have margin lengths in the range between 5 and 16, producing 10 different regions with a mean of 1107.8 voxels in each region. 

%
%


In order to demonstrate the effectiveness of the shrinkage component of the model, the true tensor coefficient values were randomly assigned using the \texttt{specifyregion} function from  \texttt{neuRosim}. This function allows us to define tensors such that nonzero elements are spatially-contiguous spheres. In this simulation, the coefficient tensors are designed such that all elements took the value of either zero or one. In real fMRI data, activation is typically observed in a small number of voxels/regions. Therefore we set the sizes of the true activated cells in our simulated data to be no greater than 5\% of the total tensor size. The true values for a slice of one of the coefficient tensors can be seen in Web Figure 2. 

The contrast-to-noise ratio, defined as $B_{g,v} / \sigma_y$ for $B_{g,v} \neq 0$ \citep{welvaert2013definition},
was set to be equal to 1, which is proposed as a realistic value for neuroimaging data by \cite{rowe2004complex}. The connectivity between tensor regions was simulated by setting two pairs of the ten regions to have a region-wide correlation of 0.9, while all other regions were assigned correlations of zero. A covariance matrix ($\boldsymbol{\Sigma}^{-1}$) was created from this correlation matrix, and the region effects for subject $i$ were simulated from a multivariate normal distribution with mean zero and covariance $\boldsymbol{\Sigma}^{-1}$. The signal-to-noise ratio, defined as $\Sigma_{i,j}^{-1} / \sigma_y^2$ was set to 5, a realistic value based on \citet{welvaert2013definition}. This quantity can be thought of as the relative effect of the connectivity on the observed response tensors. Finally, the observation-level variance ($\sigma_y^2$) was set to be 1. 
\newline
\newline \underline{\emph{Competitors}}\\
We fitted our proposed Bayesian model to the simulated data using different choices of rank $R$. In most of the real life applications, smaller values of $R$ are sufficient to attain the desired inference, hence the model was tested for ranks 1 through 5. The performance of the proposed model is compared with 
an alternative approach that vectorizes the tensor response, builds voxel specific regression model by regressing the response on predictors, followed by jointly estimating the voxel specific regression coefficients using a shrinkage prior distribution. More precisely, if $Y_{i,g,t,v_1,v_2,v_3}$ is the response at voxel $(v_1,v_2,v_3)$ in region $g$ at time $t$ for individual $i$, our competing model proposes
\begin{align}\label{eq:GDP}
Y_{i,g,t,v_1,v_2,v_3}\stackrel{ind.}{\sim} N(b_{g,v_1,v_2,v_3}^* x_{i,t}+d_{i,g}^*,\sigma^{*2}),
\bbeta_g^* \sim \text{N}(\mathbf{0},\tau_g^* \mathbf{W}_g^*),
\end{align}
where $\bbeta_g^*=(b_{g,v_1,v_2,v_3}^*: v_1=1:p_{1,g},v_2=1:p_{2,g},v_3=1:p_{3,g})' \in\mathbb{R}^{p_{1,g}\times p_{2,g}\times p_{3,g}}$ is the vector of coefficients and
$\mathbf{W}_g^*=(\omega_{g,v_1,v_2,v_3}^*: v_1=1:p_{1,g},v_2=1:p_{2,g},v_3=1:p_{3,g})$. $d_{i,g}^*$'s are jointly assigned a Gaussian graphical prior similar to (\ref{graph_lasso}). The hierarchical specification is completed by assigning $\tau_g^* \sim \text{Gamma}(a_\tau, b_\tau)$, $\omega_{g,v_1,v_2,v_3}^*  \sim \text{Exp}\left( \frac{\lambda_g^{*2}}{2} \right)$, $\lambda_g^* \sim \text{Gamma} (a_\lambda,b_\lambda)$. We coin this approach as \emph{vectorized-GDP}. Comparison with this reveals 
the advantage of retaining the tensor structure of the response, as well as the advantage due to the parsimony offered by the PARAFAC decomposition. We also attempted to implement a spatially varying coefficient (SVC) model \citep{zhang2015bayesian} and found to be computationally demanding due to large matrix inversions in each MCMC step. Hence the comparison with SVC is not reported. 
\newline
\newline \underline{\emph{Comparison metrics}}\\
MCMC is run for 1100 iterations for all competitors, with a 100 
burn-in and the remaining used for inference. The assessment of convergence is made by the Raftery-Lewis diagnostic test implemented in the \texttt{R} package ``coda". It shows a median effective sample size of 1,000 for the elements of $\mathbf{B}_g$ in the rank 1 model, decreasing to a median effective sample size of 684 in the rank 5 model after the burn in, which seems sufficient for satisfactory inference.

Comparisons among competitors are based on (a) a model fitting statistic, (b) point estimation of $\bB_g$'s and (c) frequentist coverage (and length) of 95\% credible intervals (CI). The accuracy of detecting active and inactive voxels for each competitor are also reported. Finally, we compare competitors in terms of identifying connectivity between regions.





Model fitting is compared using the deviance information criterion (DIC), defined in \citet{gelman2014bayesian} as $\text{DIC} = -2\log p(\mathbf{Y}|\hat{\mathbf{B}},\hat{\mathbf{d}},\mathbf{X},\hat{\sigma}_y^2) + 2p_{DIC},$ where\\ $p_{DIC} = 2 \left( \log p(\mathbf{Y}|\hat{\mathbf{B}},\hat{\mathbf{d}},\mathbf{X},\hat{\sigma}_y^2) - \frac{1}{S} \sum_{s = 1}^{S} \log p(\mathbf{Y}|\mathbf{B}^s,\mathbf{d}^s,\mathbf{X},\hat{\sigma}_y^{2(s)})  \right),$
$\hat{\theta}$ is the posterior mean of any parameter $\theta$, $S$ is the total number of post burn-in posterior samples. The superscript $s$ denotes $s$th post burn-in posterior sample for a parameter, $\bY$, $\bX$ are the collection of all responses and predictors respectively. In order to correct for any outlier posterior draws, the DIC was calculated by thinning the posterior sample to every four draws after burn-in.

For comparison between the models in terms of point estimation of $\bB_g$'s, we compute square root of the mean squared error between the estimated tensor coefficient and true tensor coefficient, $\sqrt{\sum_{g=1}^{G}\sum_{v\in\mathcal{R}_g}(\bar{B}_{g,v} - B_{g,v}^0)^2}$,
where $\mathcal{R}_g$ represents region $g$, $B_{g,v}^0$ and $\bar{B}_{g,v}$ are the true and the posterior mean of the $v$the cell coefficient in the $g$th region respectively.

Given the posterior mean estimates of ${B}_{g,v}$, sequential two-means approach \citep{li2017variable} is employed to identify active and inactive voxels. True positive rate (TPR) and false positive rate (FPR) are computed corresponding to different thresholds and the area under the receiving operating characteristic (ROC) curve, known as the AUC, is presented as a measure of how well the truely active and inactive voxels are detected by the proposed method. Finally, uncertainty quantification by the competitors is assessed by length and coverage 95\% posterior credible intervals. We also report computation times for the competing models.

\begin{table}
\centering
\addtolength{\leftskip} {-3cm}
\addtolength{\rightskip}{-2cm}
\caption{Performance diagnostics based on 1,100 draws from the posterior distribution with multiple different models using the same simulated data. For the DIC, RMSE, AUC, and 95\% interval length and coverage, the first 100 draws from the posterior distribution are discarded as a burn-in.}
{\scriptsize
\begin{tabular}{rrrrrrrr}
  \hline
 & \# Parameters & Time (Hrs) & log(DIC) & RMSE for B & AUC & 
    \multicolumn{1}{p{2.5cm}}{\centering 95\% Credible \\ Interval Lengths} &
    \multicolumn{1}{p{2.5cm}}{\centering 95\% Credible \\ Interval Coverage} \\
  \hline
Rank 1 & 309 & 3.13 & 21.8085 & 0.1890 & 0.5495 & 0.0298 & 0.3091 \\ 
  Rank 2 & 618 & 5.24 & 21.8094 & 0.1184 & 0.8775 & 0.0336 & 0.8176 \\ 
  Rank 3 & 927 & 6.20 & 21.8080 & 0.0807 & 0.9483 & 0.0364 & 0.8992 \\ 
  Rank 4 & 1236 & 6.93 & 21.8310 & 0.0679 & 0.9880 & 0.0415 & 0.9247 \\ 
  Rank 5 & 1545 & 7.84 & 21.8104 & 0.0681 & 0.9877 & 0.0432 & 0.9604 \\ 
  Vectorized & 11078 & 2.15 & 21.8286 & 0.1129 & 0.9096 & 0.2115 & 1.0000 \\ 
   \hline
\end{tabular}
}
\label{tableone}
\end{table}
\noindent\underline{\emph{Results}}\\
Performance measures for all the competitors are summarized in Table \ref{tableone}. The proposed model with various ranks show significant improvement in terms of model fitting statistic over the vectorized-GDP. 
The tensor models also demonstrate benefit in terms of point estimation and uncertainty quantification. While all tensor models of rank more than 3 show close to nominal coverage, vectorized-GDP suffers from over-coverage with a much wider 95\% credible interval. In fact, tensor models corresponding to Rank 3, 4, 5 show excellent detection of activated regions, as is witnessed in Figure 2 and AUC column of Table~\ref{tableone}. Although Rank 4 and 5 models demonstrate marginally lower RMSE and improved coverage over the Rank 3 model, Rank 3 model enjoys lowest model fitting statistic. Perhaps, the Rank 4 and 5 models are penalized for having a large number of parameters (see Table~\ref{tableone}).  Overall, tensor models with rank greater than $2$ comprehensively outperform vectorized-GDP as competitors.



Similar to the tensor coefficient, sequential two-means method \citep{li2017variable} is used on the off-diagonal elements of the precision matrix $\boldsymbol{\Sigma}$ to recover the connectivity structure among regions in the simulated data. 
All unconnected regions are classified to have a partial correlation of zero, and the connected regions have nonzero partial correlations. The estimates are shown in Web Figure 3. Importantly, the assessment of connectivity appears to be accurate and robust across tensor models of all ranks. The underestimation of the partial correlation has two causes: a relatively low signal-to-noise ratio, and the shrinkage of the estimates that stems from the application of a strong shrinkage prior on the off-diagonal $\boldsymbol{\Sigma}$.



\noindent\underline{\emph{Hyperparameter sensitivity}}\\
Finally, in order to test the robustness of the model to choices of the hyperparameters, a grid of hyperparameter values was made by scaling each of the ``standard'' values for $a_\lambda$, $b_\lambda$, $a_\tau$, $b_\tau$, $a_\zeta$, $b_\zeta$, $a_\sigma$, and $b_\sigma$, defined in section \ref{subsec:hyperspec}, by 0.01, 1, and 100, resulting in 6,561 different combinations. Of these, 100 settings were randomly sampled from the list and then tested with tensor model corresponding to Rank 3. We graphed boxplots of the RMSE as well as length and coverage of 95\% CI for all these hyperparameter combinations (these are available in Web Figure 4 of the Supplementary Material). 
The results are fairly robust with all three metrics varying within a small range under all different hyperparameter combinations.
Overall, the simulation study reveals excellent recovery of activation and connectivity among regions by the proposed model. Although the computation time for the proposed model may a bit on the higher side, the burden is somewhat lessened by the rapid MCMC convergence for the model parameters allowing accurate inference even with a small burn-in.

\section{Real Data Analysis}
\label{sec:real_data_analysis}

We analyze data collected in a study examining the fMRI scans of individuals undergoing a test which introduces risk-taking scenarios. This study is known as the Balloon Analog Risk-Taking Task Experiment \citep{schonberg2012decreasing}. \citep{guhaniyogispencerbayesian} also presented a brief analysis of this data, though only a single subject is chosen for the analysis. The data are available from the OpenfMRI project and the OpenNeuro platform at \url{https://openneuro.org/datasets/ds000001/versions/00006?app=MRIQC&version=33&job=5978f5dca1f52600019e85c4}. It consists of 16 individuals who were scanned using a 3T Siemens AG Allegra MRI machine in the Ahmanson-Lovelace Brain Mapping Center at UCLA. While in the scanner, the subjects inflated simulated balloons. A trial is defined as a balloon that can be pumped a certain number of times. Each trial could end in one of two ways. First, the subject could ``cash-out" at any point during the trial and add the cumulative winnings for that balloon to their collective ``bank". Second, upon pumping, a balloon may explode and the participants would lose the cumulative winnings for that balloon and nothing would be added to their collective ``bank". The subjects interacted with the simulation by pushing one of two buttons with their right pointer finger or right middle finger. Each trial began with winnings of \$0.25, displayed below the balloon, and each successive pump added \$0.25 to the cumulative winnings for that balloon. The balloons were red, green, or blue in color, and the maximum number of pumps for a balloon was drawn from a discrete uniform distribution between 1 and 8, 12, or 16, depending on the color of the balloon. Intermittently, subjects would be shown a grey control balloon with a maximum of 12 pumps that did not explode and did not have any associated monetary value. Unlike with the colored balloons, subjects did not have the option to ``cash-out" when inflating the control balloon. Each run for each subject was ten minutes in length, during which each color balloon could be presented no more than 12 times.

Using FSL \citep{smith2004advances}, the fMRI scans were smoothed to correct for motion using a Gaussian kernel with a full-width half-maximum (FWHM) of 5mm, and then mapped to the Montreal Neurological Institute (MNI) standard in order to compare scans between subjects with different brain sizes. A slice of the brain at z = -4 was then extracted from the larger response tensor for each subject. This slice was chosen because it has large contiguous regions, which should present a challenge when attempting to classify an appropriate number of active and inactive voxels in the brain. The data were separated into 9 regions of interest based on the MNI structural atlas \citep{evans1994mri}. 

\begin{figure}
    \centering
    \includegraphics[width = \textwidth]{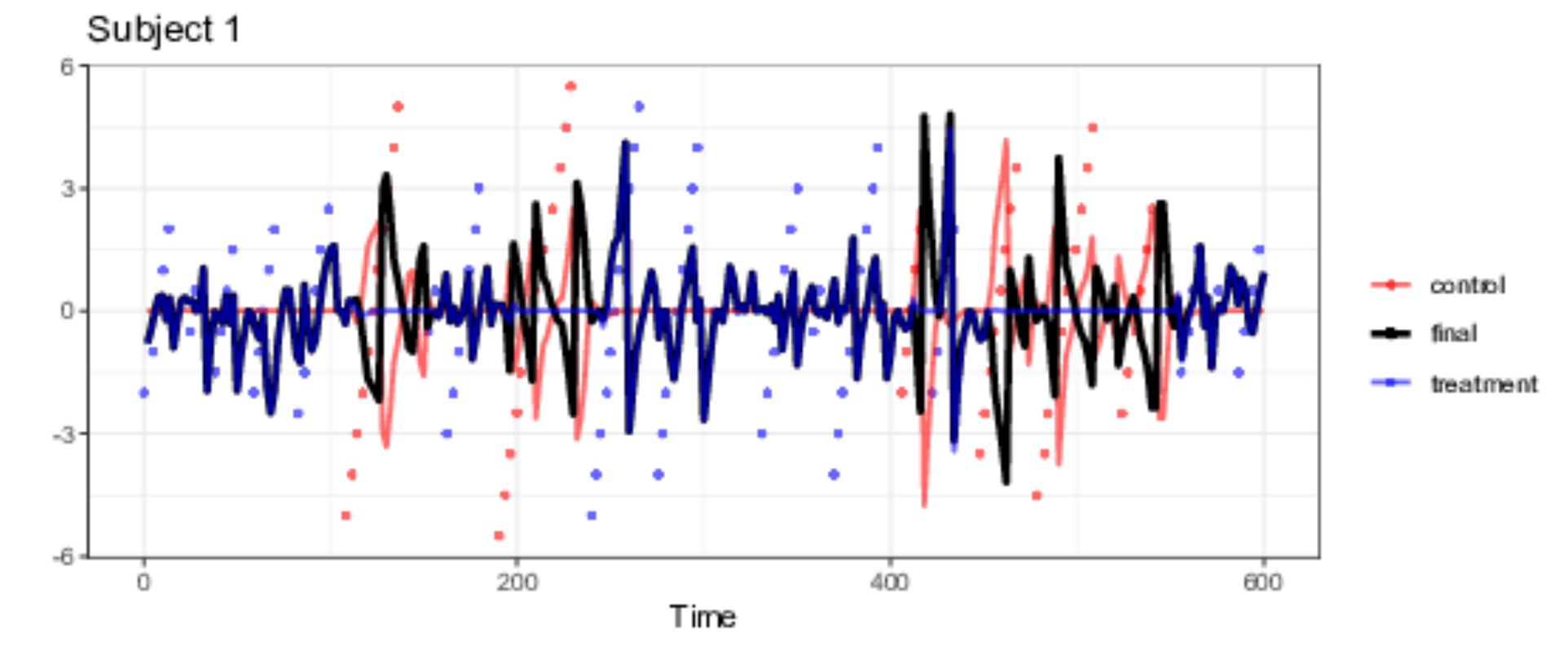}
    \caption{The numeric values for the centered number of pumps for the control and treatment balloons, their convolutions with the double-gamma haemodynamic response function, and the final covariate used for these analyses.}
    \label{fig:916_convolved_pumps}
\end{figure}

The regions vary in size, and the median number of voxels per region for each subject is 646. To measure the level of risk being processed by a subject at a given time, begin with the centered number of pumps that an individual gave a “treatment” balloon before they “cashed-out” or the balloon exploded. It is assumed that the higher the number of pumps becomes, the more risk is present to the individual. 
This value was then convolved with the double-gamma haemodynamic response function, which takes into account the physiological lag between stimulus and response, and smooths the stepwise function for the centered number of pumps. Finally, the centered, convolved number of pumps on the control balloon is subtracted from the treatment series to provide a basis of comparison. Figure \ref{fig:916_convolved_pumps} shows the raw values for the centered number of control and treatment pumps, as well as the convolved pump functions and the final values for the covariate that were used in these analyses. The haemodynamic response function was defined using the default values given in the \texttt{specifydesign} function of the \texttt{neuRosim} package in R.

%
%

The independent variable was then created as the difference between parametric modulation of the number of pumps on the treatment balloons and on the control balloons. The same covariate is used in one of the analyses done in \citet{schonberg2012decreasing} (see Figure \ref{fig:916_convolved_pumps}).

We fitted our proposed Bayesian tensor mixed effect model with ranks $R=1,2,3,4,5$ with the prior structure specified in Section 2. Similar to simulation studies, 1,100 samples were drawn from the joint posterior distribution of all of the parameters, and the first 100 samples were discarded as a ``burn-in" measure.
The \texttt{effectiveSize} function within the \texttt{coda} package in R is used to calculate median values for the effective sample size for the 1,000 posterior draws of the elements in all $\mathbf{B}_g$ for the five different rank models, see Table \ref{tab:ESSandDIC}. Table \ref{tab:ESSandDIC} indicates fairly uncorrelated post burn-in posterior samples to draw reliable posterior inference.

%
%

\begin{figure}
\begin{center}
\includegraphics[width=\linewidth]{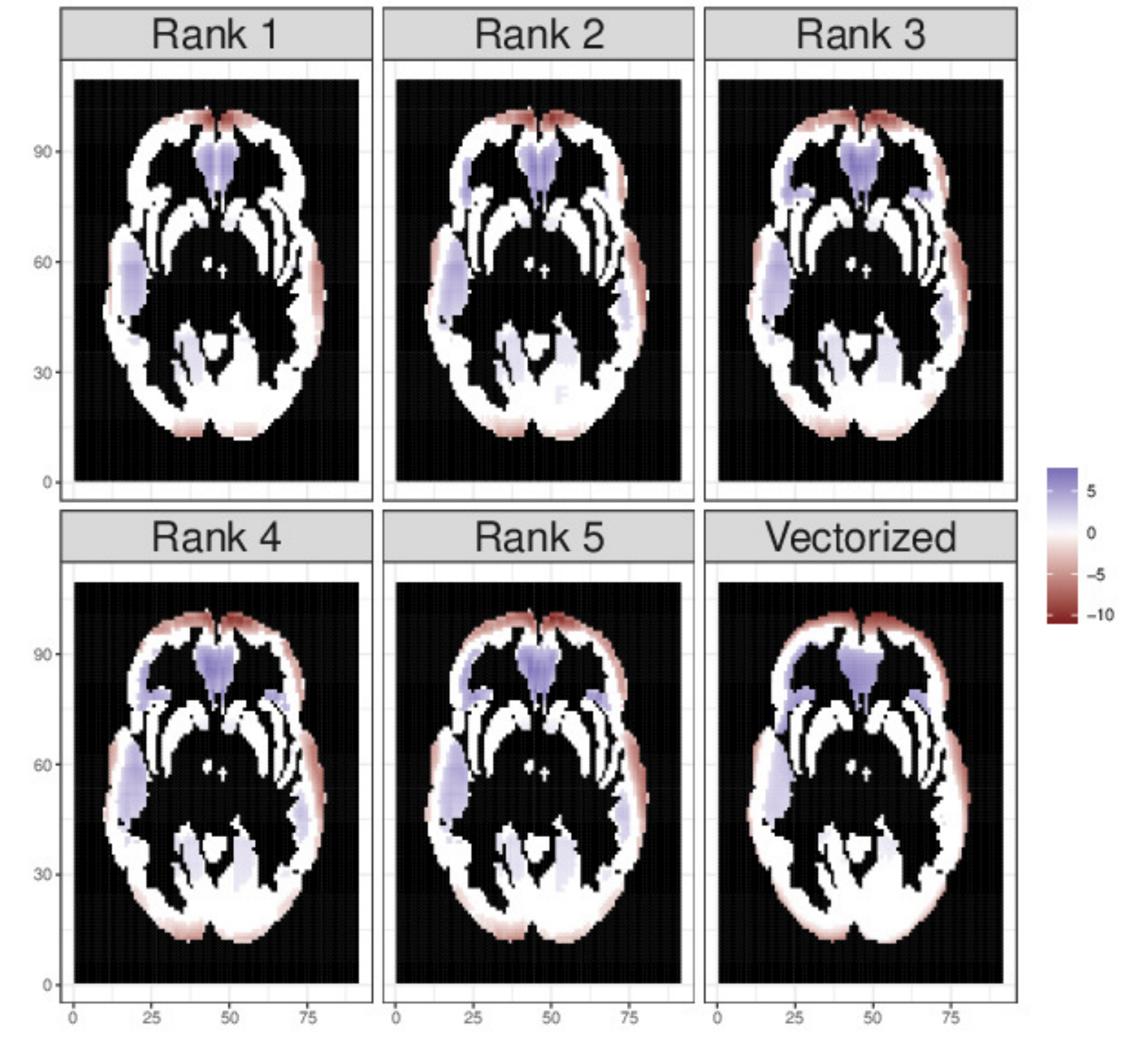}
\caption{The posterior median results for the rank 1 through 5 models after using sequential 2-means to classify coefficient values as zero or nonzero. For comparison, vectorized model estimate is also included. Black regions were not analyzed, as they were not included in any regions in the Montreal Neurological Institute Atlas.}
\label{fig:post_median}
\end{center}
\end{figure}

The posterior median tensors $\mathbf{B}_g$ within the brain have been reorganized to their original positions, and can be seen in Figure \ref{fig:post_median}. Higher values of the coefficient means that there is more blood flow associated with higher levels of perceived risk. Larger positive values indicate that blood flow increases in these regions as risk increases. Larger negative values show regions that exhibit a decrease in blood flow as risk increases, perhaps indicating that blood flows from these regions to the regions with increased blood flow. Similar to the simulation studies, the vectorized GDP model competitor is also fitted to the data to assess the advantages of preserving the tensor structure of the brain image in our proposed model.
According to the Deviance Information Criterion (DIC) \citep{gelman2014bayesian} given in Table \ref{tab:ESSandDIC}, Rank 3 is the best performing tensor mixed effect model. Importantly, Rank 3 model also yields considerably smaller DIC value than the vectorized GDP. Figure \ref{fig:post_median} shows that all the models considered generally agree in terms of the posterior activation results, however, the tensor models provide more differentiated estimates of activation strength than those obtained from the vectorized model, particularly in the Frontal Lobe.

\begin{table}[ht]
\centering
\caption{The median effective sample size and log deviance information criterion for the five rank models.}
\begin{tabular}{rrrrrrr}
  \hline
 & Rank 1 & Rank 2 & Rank 3 & Rank 4 & Rank 5 & Vectorized \\ 
  \hline
median ESS & 1000.0000 & 974.2903 & 874.0009 & 845.8368 & 845.8359 & 1000.0000 \\ 
  log(DIC) & 22.5430 & 22.5442 & 22.5428 & 22.5552 & 22.5551 & 22.5455 \\ 
   \hline
\end{tabular}
\label{tab:ESSandDIC}
\end{table}

\begin{figure}
    \centering
    \includegraphics[width=\textwidth]{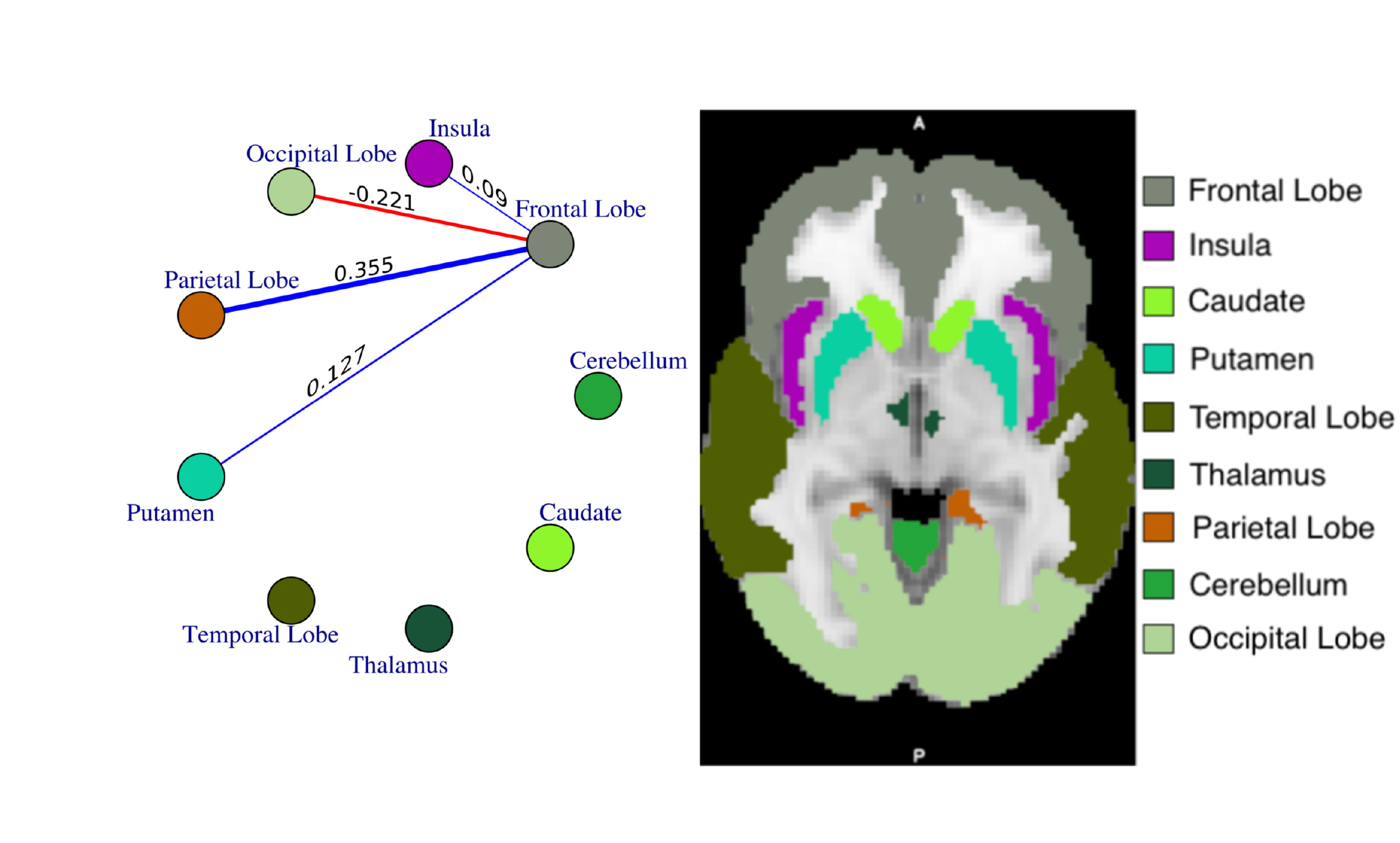}
    \caption{The connected regions of the brain in the rank 3 model, based on the partial correlation. The partial correlation here was found after  using the sequential two-means method \citep{li2017variable} on the precision matrix elements.}
    \label{fig:rank_3_partial_correlation}
\end{figure}

The estimates for the partial correlations between regions shown in Figure \ref{fig:rank_3_partial_correlation} indicate that the Frontal Lobe plays an
important role in this task showing significant positive partial correlation with the Insula, Parietal Lobe and Putamen and significant negative partial correlation with the Occipital Lobe. This agrees with earlier experiments suggesting that the frontal lobe plays a role in the assessment of risk \citep{miller1985cognitive}.



%
%

\section{Discussion}
\label{sec:discussion}



Bayesian literature on multi-subject joint modeling of voxel-level activation and ROI level connectivity is quite sparse due to the sheer volume of data from multiple subjects. Seeking to fill this gap, this article provides a Bayesian joint modeling framework for exact inference of voxel-wise brain activation and functional connectivity between predefined ROIs in fMRI data arising from multi-subject experiments. The proposed approach is based on a tensor mixed effect response model. To ensure computational flexibility as well as parsimony, PARAFAC tensor decomposition is employed for representing tensor valued activation coefficients. Additionally, Bayesian graphical modeling components are used for ascertaining connectivity between ROIs. The proposed model produces marked improvement over a vectorized-response model both in terms of identifying point estimation and quantifying uncertainty in a statistically principled manner. The robustness of the model to choices of hyperparameters, and the flexibility of the modeling structure without using basis functions makes these models accessible to a wide range neuroscientists and statisticians alike. The open source code have been written to generalize to tensors of any dimension, which may prove useful in applications outside neuroimaging.

Our proposed approach assumes several important extensions. Notably, the parsimony in activation coefficients achieved by a PARAFAC decomposition may appear to be restrictive in certain applications, and can be replaced by a more flexible Tucker decomposition. Extensions of (\ref{eg:lik_tensor}) that incorporate nonlinear regional effects through time will also be explored. Finally, investigation into model-driven choices for subject-specific haemodynamic response functions may improve upon the accuracy of the proposed approach in real data applications.

\bibliographystyle{biom}\bibliography{Advancement}

 \appendix



\section{Web Appendix A}

This posterior sampling algorithm can be done efficiently by sampling the region-specific variables in parallel.

\begin{enumerate}[(1)]
	\item Draw $\alpha_g$ via a Griddy-Gibbs algorithm as follows:
	\begin{enumerate}
		\item For each possible value of $\alpha_g$, draw a sample of size $\mathcal{M}$ from the posterior distributions of $\phi_{r,g}$ and $\tau_{g}$. 
		\item Evaluate the posterior density using each of these individual samples using the previous iteration values for all other parameters. 
		\item Average these densities together for each possible value for $\alpha_g$ in the grid, and then sample one value using the averaged densities as weights.
	\end{enumerate}
	\item Using the posterior full conditional kernel of 
	\begin{align*}
	p(\xi_{g,r}|\boldsymbol{\beta}_{g,\cdot,r},\mathbf{W}_{g,\cdot,r},\xi_{g,-r},\tau_g) \propto &  \, \xi_{g,r}^{-\sum p_j / 2} (1 - \xi_{g,r})^{-(R - r) \sum p_j /2} \\
	& \times \exp \left\{ -\frac{1}{\tau_g} \left[ \frac{1}{\xi_{g,r}} \sum_{j = 1}^{D} \left( \boldsymbol{\beta}_{g,j,r}^T \mathbf{W}_{g,j,r}^{-1} \boldsymbol{\beta}_{g,j,r} \right)  \right.  \right. \\
	& \left. \left.  + \sum_{k=r+1}^{R} \frac{1}{\xi_{g,k} \prod_{\ell = r}^{k - 1} (1 - \xi_{g,\ell}) } \sum_{j = 1}^{D} \left( \boldsymbol{\beta}_{g,j,r}^T \mathbf{W}_{g,j,r}^{-1} \boldsymbol{\beta}_{g,j,r} \right) \right] \right\}, 
	\end{align*}
	draw $\xi_{g,r}^{(s)}$ for sample $s$ using a Metropolis-Hastings step with a normal proposal distribution with mean $\xi_{g,r}^{(s-1)}$ and variance $0.01^2$. The value for the variance was chosen such that the datasets tested showed decent mixing. After drawing $\xi_{g,1}, \ldots, \xi_{g,R-1}$, set $\phi_{g,r} = \xi_{g,r} \times \prod_{k=1}^{r - 1} (1 - \xi_{g,k})$, and set $\phi_{g,R} = 1 - \sum_{r = 1}^{R - 1} \phi_{g,r}$.
	\item Draw each $\tau^g$ from a generalized inverse Gaussian distribution, $gIG(\nu,\chi,\psi)$, where
	\begin{align*}
	\nu & = a_\tau - \frac{R \sum p_j}{2}, & 
	\chi & = \sum_{r=1}^{R} \frac{1}{\phi_r}\left( \sum_{j=1}^{D} \boldsymbol{\beta}_{j}^{(r)'}\mathbf{W}_j^{(r)-1}\boldsymbol{\beta}_{j}^{(r)} \right), &
	\psi & = 2b_\tau
	\end{align*}
	
	\item Draw each $\lambda_{j}^{(r)g}$ from a 
	$$\text{Gamma}\left( a_\lambda + p_j, b_\lambda + \frac{1}{\sqrt{\phi_{r}^g \tau^g}} \sum_{\ell=1}^{p_j} \left| \beta_{j,\ell}^{(r)g} \right| \right)$$
	\item Draw each $\omega_{j,\ell}^{(r)g}$ from a generalized Inverse Gaussian distribution, $gIG(\nu,\chi,\psi)$, where
	\begin{align*}
	\nu & = \frac{1}{2} &
	\chi & = \frac{\beta_{j,\ell}^{(r)g2}}{\tau^g \phi_{r}^g} &
	\psi & = \lambda_{j}^{(r)g2}
	\end{align*}
	
	\item When $D=2$, draw each $\beta_{j,z}^{(r)g}$ from a normal distribution with variance
	
	$$ \boldsymbol{\Lambda} = \left( \frac{1}{\phi_{r}^g \tau^g \omega_{j,z}^{(r)g}} + \frac{n\sum x_t^2 \boldsymbol{\beta}_{-j}^{(r)g'}\boldsymbol{\beta}_{-j}^{(r)g}}{\sigma_y^2} \right)^{-1}$$
	
	and mean
	
	$$ \boldsymbol{\mu} = \boldsymbol{\Lambda} \frac{  \sum \sum x_t \boldsymbol{\beta}_{-j}^{(r)g'} \hat{\mathbf{y}}_{t,i}^g  }{\sigma_y^2}$$
	
	where 
	$$\hat{\mathbf{y}}_{t,i}^g = \mathbf{y}_{t,i}^g - d_i^g \mathbf{1} - x_t \sum_{\ell \neq r} \boldsymbol{\beta}_{1}^{(\ell)g} \circ \boldsymbol{\beta}_{2}^{(\ell)g} $$
	\item Draw each $\mathbf{d}_i$ from a normal distribution 
	$$\begin{pmatrix} d_{i,1} \\ \vdots \\ d_{i,G} \end{pmatrix} = \mathbf{d}_i \sim \text{N}(\boldsymbol{\theta}_i,\mathbf{M})$$
	where
	$$\mathbf{M} = \left(\boldsymbol{\Sigma} + \frac{T\mathbf{V}}{\sigma_y^2} \right)^{-1}$$
	and
	$$ \boldsymbol{\theta}_i = \mathbf{M}\left( \frac{\sum_v \sum_t \tilde{\mathbf{y}}_{i,v,t}}{\sigma_y^2} \right) $$
	where $T$ is the number of time steps in the fMRI scan, $\mathbf{V}$ is a diagonal matrix where $v_{ii}$ is equal to the number of voxels in region $i$, and
	$$\begin{aligned} 
	\tilde{\mathbf{y}}_{i,v,t} & = \begin{pmatrix} 
	\tilde{\mathbf{y}}_{i,1} \\ \vdots \\ \tilde{y}_{i,G} 
	\end{pmatrix}, & \tilde{\mathbf{y}}_{i,g} = \mathbf{y}_{i,g} - \mathbf{B}\mathbf{x} 
	\end{aligned}$$
	\item For each region $g$, draw $\delta_g$ from a $\text{gamma}\left( \frac{n}{2} + 1, \frac{S_{gg} + \zeta}{2} \right)$
	\item Draw $\boldsymbol{\eta}$ from a multivariate normal distribution with covariance 
	$$\boldsymbol{\varphi} = \left( (S_{gg} + \zeta) \boldsymbol{\Sigma}_{-g,-g}^{-1} + \text{diag}(1 / \boldsymbol{\Upsilon}_{-g,g}) \right)^{-1}$$
	and mean
	$$\mathbf{L} = -\boldsymbol{\varphi}\mathbf{S}_{g,-g}$$
	Set $\boldsymbol{\Sigma}_{g,-g} = \boldsymbol{\Sigma}_{-g,g} = \boldsymbol{\eta}$ and $\boldsymbol{\Sigma}_{g,g} = \delta_g + \boldsymbol{\eta}^T \boldsymbol{\Sigma}_{-g,-g}^{-1} \boldsymbol{\eta}$
	
	\item For $i > g$, draw $u_{g,i}$ from an inverse Gaussian distribution with mean $\sqrt{\left(\frac{\zeta^2}{\boldsymbol{\Sigma}_{g,i}} \right)}$ and shape $\zeta^2$. Set $\boldsymbol{\Upsilon}_{g,i} = \boldsymbol{\Upsilon}_{i,g} = 1/u_{g,i}$.
	\item Draw $\zeta$ from a $\text{gamma}(a,b)$ distribution where
	$$\begin{aligned}
	a & = a_\zeta + \frac{G(G+1)}{2} & b & = b_\zeta + \frac{\sum_i \sum_j |\Sigma_{ij}|}{2}
	\end{aligned}$$
\end{enumerate}

Following this algorithm, the MCMC converges rapidly to the region of the maximum likelihood estimator. 

\newpage

\section{Web Figure 1}

\begin{figure}[h]
	\centering
	\includegraphics[width=3in]{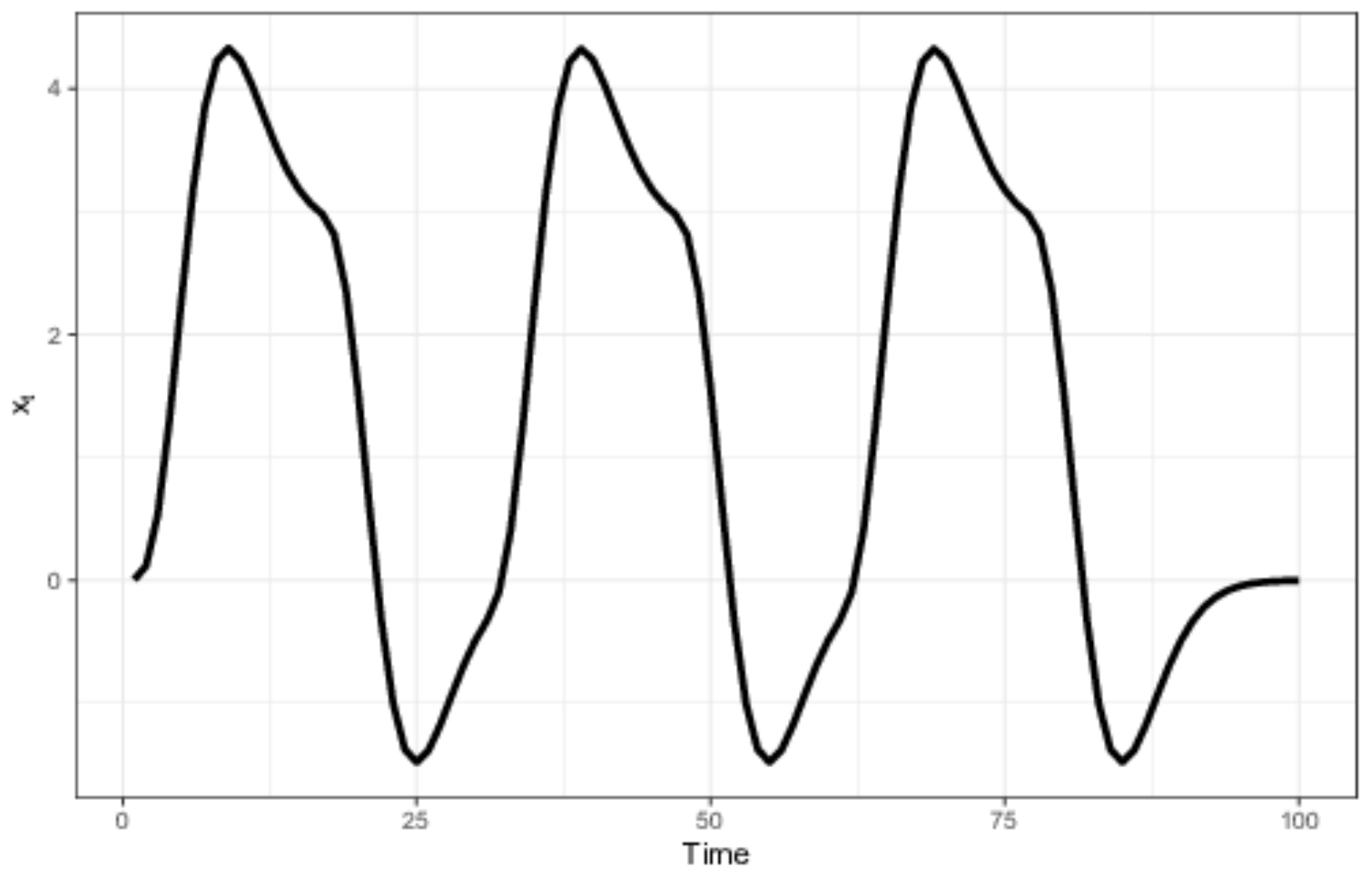}
	\caption{Values for the covariate $x_t$ in the simulated data.}
	\label{fig:sim_data_covariate}
\end{figure}

\newpage

\section{Web Figure 2}

\begin{figure}[h]
	\centering
	\includegraphics[width=\textwidth]{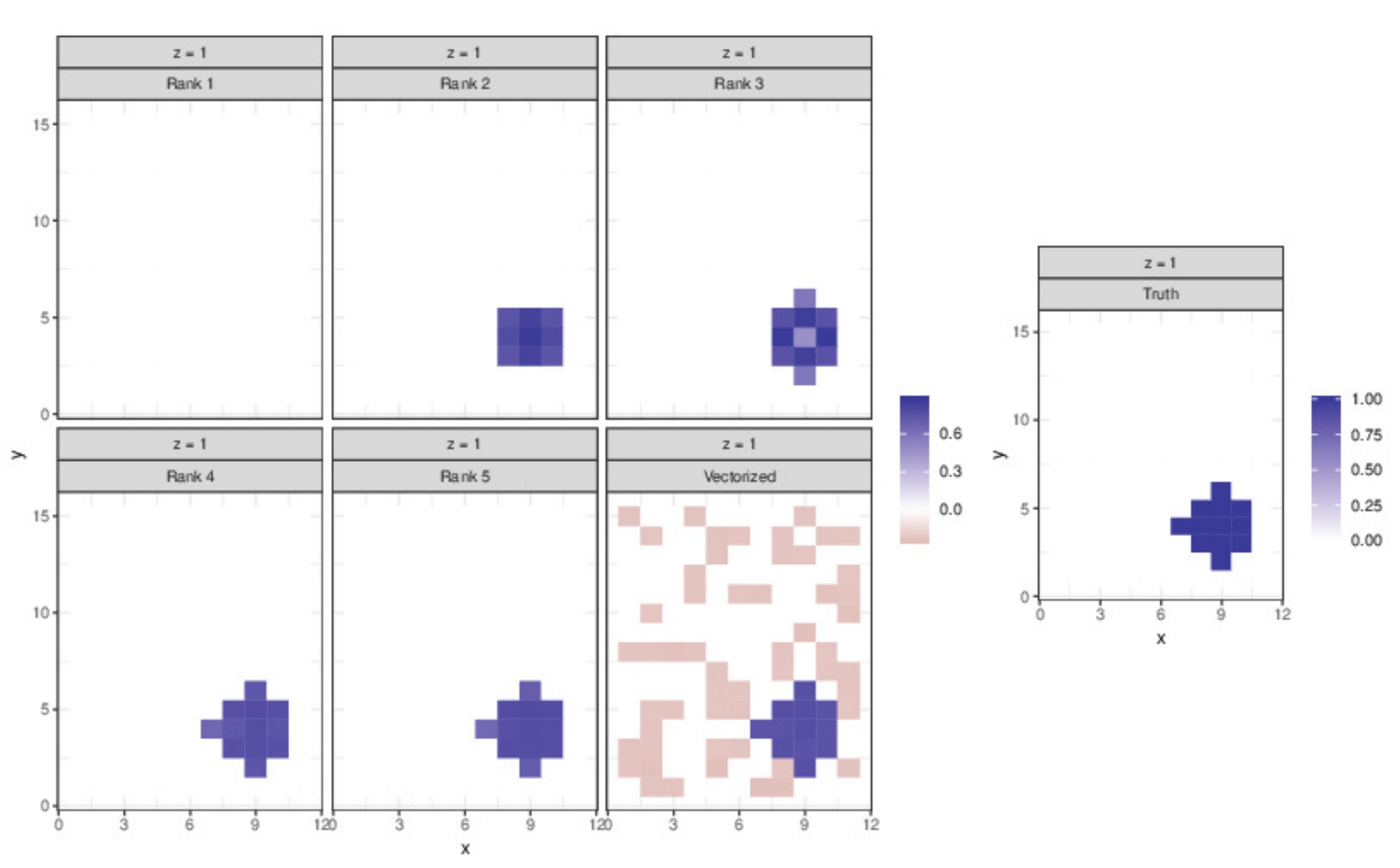}
	\caption{Rank model estimates and true value for a single slice of a three-dimensional coefficient tensor. Estimates are found using the sequential two-means variable selection method \citep{li2017variable}.}
	\label{fig:sim_activation}
\end{figure}

\newpage

\section{Web Figure 3}

\begin{figure}[h]
	\centering
	\includegraphics[width=\textwidth]{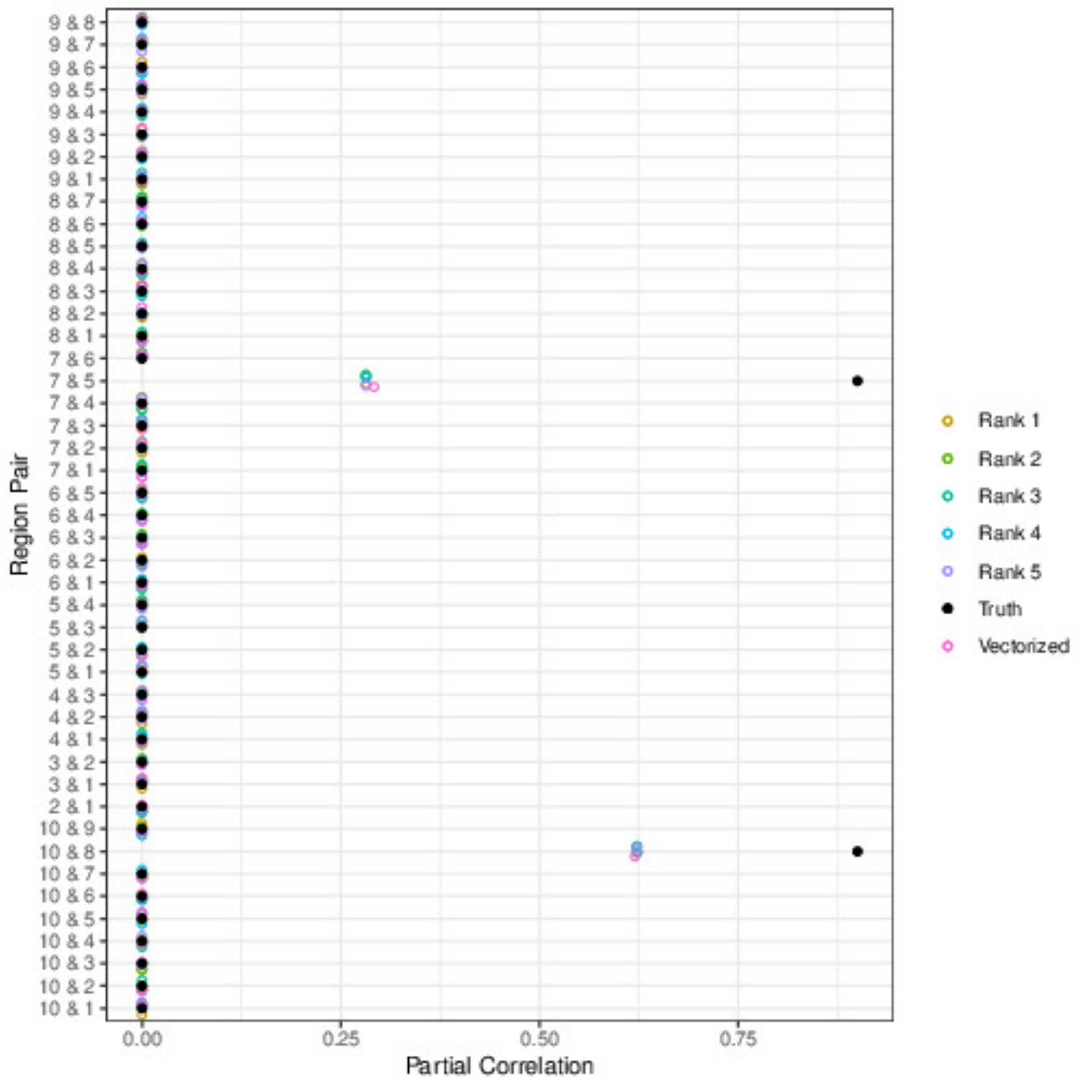}
	\caption{Estimates of the partial correlation for all possible region pairs after using the sequential 2-means method from \citet{li2017variable}. The true partial correlation values for all region pairs are shown for comparison.}
	\label{fig:sim_connectivity_plot}
\end{figure}

\newpage

\section{Web Figure 4}

\begin{figure}[h]
	\centering
	\includegraphics[width=0.7\textwidth]{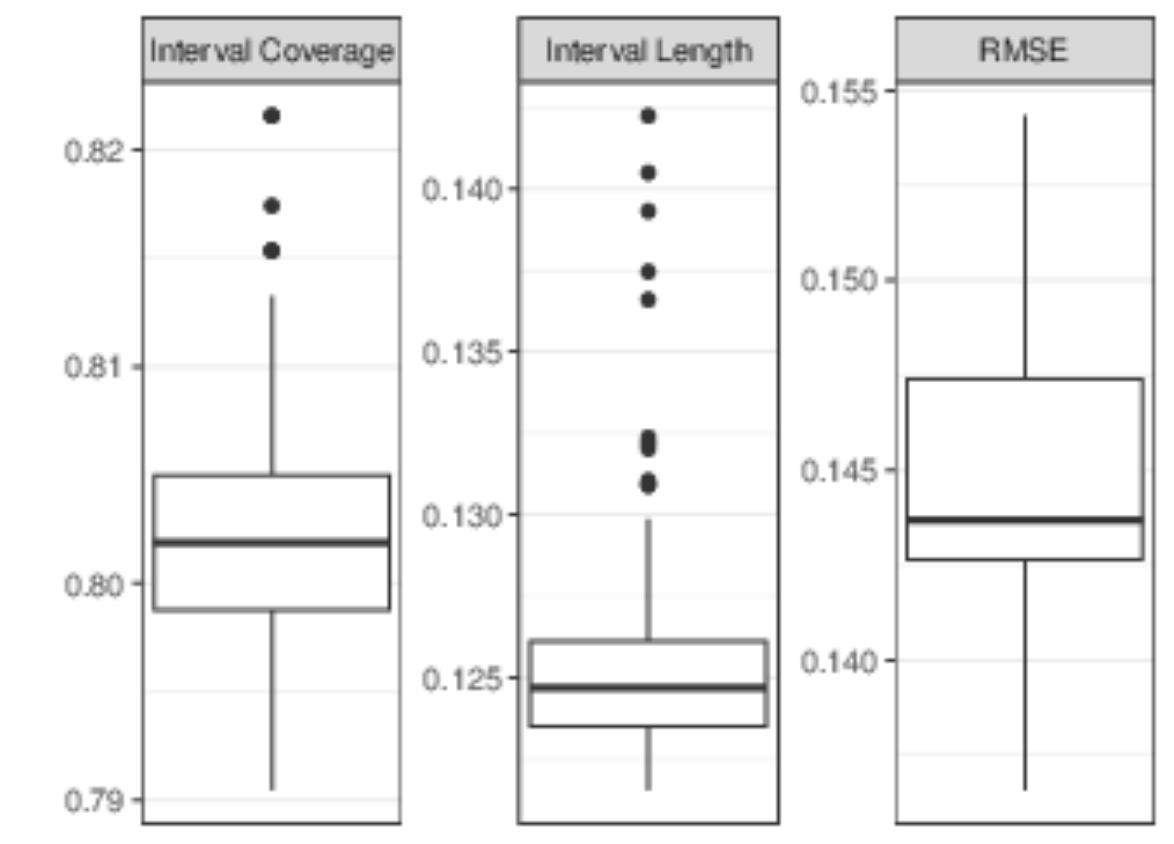}
	\caption{Boxplots for the 95\% interval coverage, interval length, and square root of the mean squared error for the 100 randomly selected hyperparameter settings.}
	\label{fig:800_hyperparameter_boxplots}
\end{figure}

\end{document}